# Quantum theory can be collectively verified

Arindam Mitra

Anushakti Abasan, Uttar Phalguni -7, 1/AF, Salt Lake,

Kolkata, West Bengal, 700064, India.

Abstract: No theory of physics has been collectively scientifically verified in an experiment so far. It is pointed out that probabilistic structure of quantum theory can be collectively scientifically verified in an experiment. It is also argued that experimentalist's point of view quantum theory is a complete theory.

Generally, a proposed theory of physics is accepted by scientific community when it is experimentally verified. It can be expected that in principle a correct physical theory can be collectively as well as individually verified in an experiment. Otherwise the proposed physical theory cannot be called to be a scientific theory at all.

It is well-known that Einstein was occupied with the idea of completeness of physical theory. Basically, the issue of completeness of physical theory initiated the great scientific debate [1,2] which is still continuing. Experimentalist's point of view a physical theory may be considered to be a complete theory if the theory can be both individually as well as collectively verified in experiment.

In the last century quantum theory, special and general theory of relativity have been repeatedly collectively verified without any ambiguity. But none of the collective verifications can be called scientific verification. The reason is, in the existing group work none of the members can scientifically rule out the possibility of manipulation of experimental data by other group members. Therefore, members of an experimental group cannot scientifically collectively claim these theories as correct theories.

It is well-known that despite his own contribution to the initial development of quantum theory Einstein could not accept its probabilistic structure. "*God does not play dice*" is his famous comment on quantum theory. Can't God(s) play dice for the verification of quantum theory ? Einstein never asked this question. The essence of quantum theory will be individually as well as collectively verified if its probabilistic structure -intrinsic randomness - is individually as well as collectively verified. Collective verification of probabilistic structure of quantum theory means parties have to unanimously accept the outcome of quantum measurement.

From the above discussion it can be understood that quantum theory can be collectively verified if uncontrollable/trustworthy random data can be collectively generated by quantum measurement. The issue of individual verification will be discussed later. It is still not known how to collectively generate uncontrollable/trustworthy random data even by classical means. This is considered to be a game-theoretic/cryptographic problem [3,4]. With the advent of quantum information the issue has been studied from that angle. It has been claimed [5] that uncontrollable random numbers/bits cannot be collectively generated within quantum theory even if noise is not considered. Interestingly, the proof allows collective generation of partly controllable non-random numbers within quantum theory. The claim was based on another claim [6,7]. As information processing is nothing but an experiment, so the claim [5] ultimately implies that quantum theory cannot be collectively verified in experiment. It means quantum theory cannot be called to be a scientific theory. It gives an extremely uneasy feeling, although many will take the conclusion as the most objectionable revelation in modern times.

In the proofs [5-7] it is implicitly assumed [8-9] that bit value is encoded in a qubit – a two state quantum system. We have noticed [10] that existing quantum cryptosystems [8], based on qubit model of information processing, cannot provide security in communication without the support of classical cryptosystem. This conceptual problem can be overcome following alternative quantum coding (AQC) technique [10,11] wherein a sequence-ensemble of quantum states represents a bit. The proofs [5-7] have no binding on AQC. Following AQC uncontrollable random data can be generated in an information processing experiment.

Suppose there is a stock of the following four EPR states of spin -1/2 particles.

$$|\psi_\pm\rangle = \frac{1}{\sqrt{2}}\left(|\uparrow\rangle\,|\downarrow\rangle \pm |\downarrow\rangle\,|\uparrow\rangle\right)$$

$$|\varphi_\pm\rangle = \frac{1}{\sqrt{2}}\left(|\uparrow\rangle\,|\uparrow\rangle \pm |\downarrow\rangle\,|\downarrow\rangle\right)$$

Choosing the four EPR states at random the pairs XX can be arranged in two rows either in direct order or in reverse order. As for example, two pairs of arrangements of 20 EPR pairs are given below.

$$S_i^0 = \{A, B, C, D, E, F, G, H, I, J, K, L, M, N, O, P, Q, R, S, T\}$$
$$S_j^0 = \{A, B, C, D, E, F, G, H, I, J, K, L, M, N, O, P, Q, R, S, T\}$$

$$S_i^1 = \{A, B, C, D, E, F, G, H, I, J, K, L, M, N, O, P, Q, R, S, T\}$$
$$S_j^1 = \{T, S, R, Q, P, O, N, M, M, L, K, J, I, H, G, F, E, D, C, B, A\}$$

The same pair of letters denotes an EPR pair. The first pair of entangled sequences $S_i^0$ and $S_j^0$, represent bit 0 and the second pair $S_i^1$ and $S_j^1$ represent bit 1. It can be seen that density matrices associated with bit 0 and 1 are same; $\rho^0 = \rho^1 = \frac{1}{4}\mathbf{I}$.

Suppose two parties, called Alice and Bob, know the basic quantum theory. Both Alice and Bob will jointly choose the EPR states to arrange them in two rows. We shall see that uncontrollable random data can be generated through bit commitment. So Alice will decide which pair of entangled sequences she will commit.

Let us describe the experimental steps with some clarifications.

**1.** Alice collects N singlets $X_i^A X_i^B$ where N need not to be large number (say, N = 50). If she collects singlets she verifies the rotational symmetry of the collected states choosing some states at random.

**2.** Alice applies unitary operators $U_i^A \in \{\sigma_x, \sigma_y, \sigma_z, I\}$ at random on each of her remaining particles $X_i^A$ where $\sigma_x, \sigma_y, \sigma_z$ are Pauli matrices and I is $2\times 2$ identity matrix. The ensemble may be described [12] by $\rho = \frac{1}{4}I$ where I is $4\times 4$ identity matrix. Alice transmits the particles $X_i^B$ to Bob and stores the partners $X_i^A$ in her quantum memory.

**3.** To verify the rotational symmetry of some of the shared states [12] Bob selects some particles $X_i^B$ at random and requests Alice to send the partners $X_i^A$ of his selected particles $X_i^B$ converting the selected pairs $X_i^A X_i^B$ into singlets. He keeps aside the remaining n particles (say n = 20) on quantum computer.

**4.** Alice applies the previously applied $U_i^A \in \{\sigma_x, \sigma_y, \sigma_z, I\}$ on partners $X_i^A$ of Bob's chosen particles $X_i^B$ to convert the chosen pairs $X_i^A X_i^B$ into singlets since $\sigma_x^2 = \sigma_y^2 = \sigma_z^2 = I^2 = I$. After this she sends the partners $X_i^A$ of the chosen pairs $X_i^A X_i^B$ to Bob.

**5.** Bob measures spin component of the pairs $X_i^A X_i^B$ along a fixed axis or randomly chosen axes. If Bob gets 100% anti-correlated data he proceeds for the next step.

**6.** Bob applies unitary operators $U_i^B \in \{\sigma_x, \sigma_y, \sigma_z, I\}$ at random on the remaining n particles $X_i^B$. Thus Alice and Bob jointly choose the four EPR states at random to arrange them in two rows.

**7.** Alice measures the spin component of her remaining n particles $X_i^A$ along z axis.

**8.** In this information processing experiment as an input Alice commits a bit 0 or 1 with probability 1/2. So Alice's input is basically a string of random bits. To commit bit 0, Alice reveals results in direct order ($D_i$). To commit bit 1, Alice reveals results in reverse order ($D_{n-i}$).

Revealing results in direct or reverse order is tantamount to arranging EPR states in direct or reverse order.

9. As an input Bob guesses the bit and reveals his guess-bit to Alice. So Bob's input is basically a string of random bits.

10. To reveal bit value Alice discloses the information of her $U_i^A$ always in direct order.

11. Firstly, Bob measures the spin components of his particles along z axis. Secondly, from the available information of $U_i^A$ and $U_i^B$ Bob reconstructs the positions of the final EPR states in two rows. Bob matches their data with the reconstructed arrangements. If Bob gets 100% EPR data in direct order ($D_iD_i$) the bit committed is 0. But if Bob gets 100% EPR data in reverse order ($D_iD_{n-i}$) the bit committed is 1.

12. If Bob's guess is right, the protocol would generate a particular bit. If wrong it will generate the other bit. Suppose 11 = 1, 00 = 1, 10 = 0, 01 = 0. Note that $2^2$ input combinations are possible. So $2^{2-1}$ input combinations can output a particular predetermined bit. Thus in this information processing experiment two input strings of random bits generates an output string of random bits. Bob will pronounce the output string to Alice to confirm that he has indeed recovered the output string.

13. Alice and bob will verify/test the randomness of the output string.

14. After the verifications of shared entanglement and the randomness of the output string they are scientifically bound to accept that they have verified quantum theory due to the following proof.

Let us prove that according to quantum theory the probability of jointly generating an uncontrollable bit by Alice and Bob is 1/2. In other words, bias given by Alice and Bob is zero; $\varepsilon_A = \varepsilon_B = 0$.

***Proof:*** It is easy to see that it is absolutely impossible for Alice to change the bit value after making the commitment and for Bob to know the bit value until and unless Alice unveils the commitment.

Density matrices of the equal mixture of four EPR-Bell states and equal mixture of the direct product states $|\uparrow\uparrow\rangle, |\uparrow\downarrow\rangle, |\downarrow\uparrow\rangle$ and $|\downarrow\downarrow\rangle$ are same.

$$\rho = \frac{1}{4}\left(|\psi_-\rangle\langle\psi_-| + |\psi_+\rangle\langle\psi_+| + |\varphi_-\rangle\langle\varphi_-| + |\varphi_+\rangle\langle\varphi_+|\right) = \frac{1}{4}\mathbf{I}$$

$$\rho = \frac{1}{4}\left(|\uparrow\downarrow\rangle\langle\uparrow\downarrow| + |\downarrow\uparrow\rangle\langle\downarrow\uparrow| + |\uparrow\uparrow\rangle\langle\uparrow\uparrow| + |\downarrow\downarrow\rangle\langle\downarrow\downarrow|\right) = \frac{1}{4}\mathbf{I}$$

where I is $4\times 4$ identity matrix. These two equations imply that EPR correlation would be completely suppressed in $\rho = \frac{1}{4}\mathbf{I}$. Here EPR correlation is independently suppressed by Alice and Bob.

Bit value can be changed if Alice can reverse the order of EPR correlation. This is possible if she can control or know Bob's data. But these two equations imply [13] that Bob's data cannot be controlled and will remain unknown to Alice because she never knows how Bob prepared $\rho = \frac{1}{4}\mathbf{I}$. Therefore, it is impossible for Alice to change the bit committed. These two equations also imply that Bob could not correlate his data with Alice's data until and unless Alice discloses how she prepared $\rho = \frac{1}{4}\mathbf{I}$ It is impossible for Bob to recover the bit committed from uncorrelated data.

Bob can guess the bit by guessing the data with probability less than 1/2. This is bad guess-work. Bob can directly guess the bit with probability 1/2 So without revealing the bit value Alice can change the bit committed with optimal probability 1/2 because it that case Bob has to directly guess the bit value.

Bob's probability of generating a bit 1 or 0 will be equal to Bob's optimal probability of knowing the bit value before its disclosure. So, $p_B = \frac{1}{2}$ and $\varepsilon_B = 0$. Alice's probability of generating a bit 1 or 0 will be equal to Alice's optimal probability of changing the bit committed. So, $p_A = \frac{1}{2}$ and $\varepsilon_A = 0$. This completes the proof.

Let us consider environmental noise. Due to noise unitary machine cannot be perfect. With imperfect unitary machine it is impossible to prepare $\rho = \frac{1}{4}I$ following the above procedure. If $\rho$ is somehow prepared then the second problem is, using advanced secret technology one can bring down the accepted noise level to unlock the suppressed entanglement. The problems can be easily overcome.

Here the advantage is, $\rho = \frac{1}{4}I$ can be prepared by simply introducing noise. After applying $U_i^A$ Alice will introduce noise until and unless entanglement is completely suppressed. After checking it Alice will send particles to Bob. Similarly, after applying $U_i^B$ Bob will introduce noise until and unless entanglement is completely suppressed. To check it Bob has to request Alice to send some particles. Thus noise can be used to overcome the disadvantage caused by noise. Next, to prevent the manipulation of environmental noise, Alice and Bob has to independently additionally introduce the noise amounting to the accepted noise level before doing anything on the shared pairs. Thus noise can be used to prevent the manipulation of noise.

It may be pointed out that by revealing wrong information error can be introduced so that data is not permanently corrupted. If correct information is later provided recovered bit can be further recovered with greater statistical confidence level.

Following the above procedure, arbitrary number of experimentalists $E_1, E_2, E_3, \ldots E_m$, stationed at m secure locations, can collectively verify the probabilistic structure of quantum theory. In this case, $E_1$ will give particles to $E_2$ who will keep some particles and transmit the remaining particles to $E_3$ without changing the order of his particles and so on. That is, each receiver will verify shared entanglement and then apply his own $U \in \{\sigma_x, \sigma_y, \sigma_z, I\}$ to prepare $\rho = \frac{1}{4}I$ ; and then each receiver will keep some particles for measurement and transmit the remaining particles to the next receiver without breaking the order of the particles. For the verification of entanglement each receiver has to choose some states which need to be converted into singlets. To do so, the preceding senders have to apply the same U which they earlier applied on the receiver's chosen state. Thus each chosen state will be converted into singlet. After this, first $E_1$ has to reveal $U_i^{E_1}$ then $E_2$ has to reveal $U_i^{E_2}$ and so on. Bit value can be recovered from the order of correlation of $E_1$'s data with one's own data. Everybody has to test randomness of the output sequence.

$E_1$ will commit a bit towards every receiver and each party except $E_1$ has to guess the bit which can be considered as input. In the output which bit 0 or 1 will be generated that depends on which guess bits are given in the inputs. Note that $2^{2m}$ input combinations are possible. So $2^{2m-1}$ input combinations will output a 0 and the rest $2^{2m-1}$ input combinations will output a 1. Uncontrollable random bits will be generated in the output because one's personal U cannot be known by other as discussed in the proof. According to quantum theory the probability of collectively generating an uncontrollable bit by m parties in the above described experiment is 1/2. Therefore, $\varepsilon_{E_1} = \varepsilon_{E_2} = \varepsilon_{E_3} = .... = 0$.

So far we discussed the possibility of verification of quantum theory generating two probable numbers 0 and 1. In the same way, arbitrary number of experimentalists stationed at m secure locations can collectively verify quantum theory generating m probable numbers. In this case instead of using two pairs of entangled sequences they have to use m pairs of entangled sequences of n EPR pairs to represent m probable numbers. All the parties have to apply $U \in \{\sigma_x, \sigma_y, \sigma_z, I\}$ to prepare $\rho = \frac{1}{4}I$ A will commit a number to all the m-1 receivers who will guess the numbers. Total $m^m$ input combinations are possible and $m^{m-1}$ different combinations of inputs can output a particular predetermined number. The probability of collectively generating an uncontrollable number by m parties is 1/m. The rest of the experimental procedures will be same as discussed above. All the parties have to tackle noise as described above. So six parties can verify quantum randomness playing dice. In other words, six Gods can play dice to verify quantum theory collectively.

Intrinsic randomness of quantum theory can be individually verified in ideal condition because in ideal condition everything can be assumed to be perfect. But in case of noisy environment state cannot be totally pure. Here the problem is, which quantum state is impure that cannot be known due to no-cloning principle [13-15]. Still, by entanglement purification [16] procedure it is possible to produce pure state in the asymptotic limit. It means intrinsic randomness can be individually verified in the asymptotic limit in non-ideal case. Apart from intrinsic randomness quantum theory can be always individually verified in other experiments.

Quantum theory is not needed to generate uncontrollable random data. Random data can be individually generated by classical means. Recently we have observed that human brain itself is a true random number generator [17,18]. In all these cases random data will be always trustworthy because classical state can be always individually verified. That is why classical theory of physics can be individually verified. The work suggests that collective verification requires non-locality. Local collective verification of physical theory cannot be trustworthy to all group members because one can manipulate data.

Einstein, Podolsky and Rosen defined [1] completeness of a physical theory from their viewpoint. They questioned [1] whether quantum theory can be called complete theory as it does not obey local realism. We have seen that due to non-locality quantum theory can be called complete theory from another view point. If this view point is accepted then the work suggests that no other theory of physics can be as complete as quantum theory.

*email:mitra1in@yahoo.com